# Longitudinal photoacoustic monitoring of collagen evolution modulated by cancer-associated fibroblasts: simulation and experiment studies

Jiayan Li
Institute of Acoustics, School of Physics Science and Engineering
Tongji University
Shanghai, China
1910791@tongji.edu.cn

Lu Bai
Department of Ultrasonography, Fudan University Shanghai Cancer Center
Shanghai Medical College, Fudan University
Shanghai, China
22211230046@m.fudan.edu.cn

Junmei Cao
Institute of Acoustics, School of Physics Science and Engineering
Tongji University
Shanghai, China
1910756@tongji.edu.cn

Wenxiang Zhi
Department of Ultrasonography, Fudan University Shanghai Cancer Center
Shanghai Medical College, Fudan University
Shanghai, China
zwenx1123@163.com

Qian Cheng
Institute of Acoustics, School of Physics Science and Engineering
Tongji University
Shanghai, China
q.cheng@tongji.edu.cn

*Abstract*—Noninvasive *in vivo* detection of collagen facilitates the investigation of mechanisms by which cancer-associated fibroblast (CAF) regulates the extracellular matrix. This study explored the feasibility of photoacoustic spectrum analysis (PASA) in identifying longitudinal changes of collagen modulated by CAFs using simulations and experiment studies. Optical and acoustic simulations in tissues were performed based on the histological slides of maximum cross-sections of murine malignancies to verify the effectiveness of photoacoustic (PA) detection system and the parameter "relative area of power spectrum density (APSD)". Experiments were conducted on three groups of mouse models with incremental ratios of CAFs and breast cancer cells at 3 continuous time points. Results discovered that the system configuration and APSD were capable of reflecting the evolution of collagen during cancer growth. Furthermore, cancers receiving a high dose of CAFs exhibited a suppressed collagen level. The presented methods show great potential for clinical translation of PASA in the field of cancer therapies targeting CAFs.

*Keywords—collagen, photoacoustic spectral analysis, cancer-associated fibroblasts, breast cancer, extracellular matrix*

## I. INTRODUCTION

Cancer-associated fibroblast (CAF) is a major component of cancer stroma and performs various functions. Activation of CAFs occurs during cancer progression through multiple mechanisms, transiting them from a quiescent state. Activated CAFs can reorganize collagen fibers within the extracellular matrix (ECM), which maintains the homeostasis of the tumor microenvironment (TME). This process results in an elevation in tissue stiffness, leading to the growth of vasculature with increased permeability and accelerating the proliferation of cancer cells. Ongoing research is being conducted to create therapeutics that specifically target CAFs by investigating the clinical outcomes related to the quantity and impact of CAFs. CAF-modulated ECM synthesis and remodeling can be detected *in vivo* using second harmonic generation imaging microscopy (SHIM) and shear wave imaging (SWI). However, the capabilities of SHIM and SWI are constrained by their shallow detection depth and lack of direct validation.

The photoacoustic (PA) modality, which relies on the emission of ultrasonic signals by chromophores when subjected to pulsed light, holds significant potential for evaluating ECM. PA technology is predominantly dependent on the light absorption properties of biomacromolecules rather than on the optical phase, which explains its superior depth of detection compared to most optical detection methods. Photoacoustic spectral analysis (PASA) can characterize the molecular composition and cluster size in biological tissues. Wang et al. successfully diagnosed fatty and cirrhotic liver by examining the slope and midband-fit of the PA power spectrum [1]. Feng et al. quantitatively analyzed the bone microstructure and marrow chemical compositions by computing the weighted center frequency and power of PA signals [2]. As far as we know, only a few existing studies achieve long-term quantitative monitoring of breast cancer using PASA. Nevertheless, these researches did not establish a connection between the PA spectral features and molecular properties.

Numerical modelling and simulation technologies greatly contribute to designing and confirming the effectiveness of innovative biomedical detection methods. They circumvent challenges caused by acquiring authentic biological specimens and ethical approvals. PA simulation entails two physical processes: the forwards propagation of light and the backwards spreading of ultrasound waves. The Monte Carlo method is widely regarded as the "gold standard" for light diffusion modelling. It considers photon scattering and absorption as random events and establishes probability distributions for the scattering angle and step size based on the optical characteristics of tissues. Besides, the k-Wave software is commonly employed in ultrasonic simulation of PA procedures due to its high computational efficiency, optimized memory requirements, and straightforward modelling techniques. It enables accurate and rapid computation of pressure fields in the time domain without being constrained by the geometry of acoustic sources. Yao et al. utilized a 3D hybrid model based on the anatomical and vascular knowledge of mouse brain to replicate the PA computed tomography [3]. Li et al. optimized a portable breast PA imaging system using 3D models derived from B-scan ultrasound data of patients [4]. Therefore, the numerical simulation holds great potential for evaluating the performance of PASA approaches in monitoring the growth of collagen induced by carcinogenesis.

This study explores the feasibility of PASA in longitudinal monitoring collagen evolution in murine breast cancers when treated with CAFs. Simulations were performed to investigate the distribution of light fluence in tissues and the propagation of PA signals based on models built from histological slides. Experiments were conducted on three groups of nude mouse cancers cultured from an incremental ratio of CAFs and cancer cells once the cancers reached average sizes of 8, 12, and 16 mm.

## II. MATERIALS AND METHODS

### A. Animal models

Female BALB/c nude mice aged four-weeks were randomly allocated into three equal groups of five mice each. $1.5\times10^6$ MDA-MB-231 cells were combined with 0, $1.5\times10^6$, $3.0\times10^6$ CAFs in a 200 μL 1:1 mixture of phosphate buffer solution with Matrigel. These three different cell blends were injected into the mammary glands of murine models. *In vivo* PA measurements were taken once the cancers reached 8 mm in diameter, defined as Day 1. Measurements were repeated on Days 8 and 18, when the average cancer diameters were 12 mm and 16 mm. Some mice were excluded from the experimental groups due to insufficient growth rate of cancer. Pentobarbital sodium (1%) was injected into the abdominal cavity of mice to induce unconsciousness during the testing process, which lasted around 30 minutes per animal. The study adhered to the regulations set by the Department of Laboratory Animal Science at Fudan University.

### B. In vivo PA experimental setup

**Figure 1(a)** illustrates the *in vivo* PA experimental system. A pulsed laser of 1580 nm emitted by an optical parametric oscillator (Phocus Mobile, OPOTEK, Carlsbad, CA) was divided into two segments with an optical energy ratio of 1:9 and transformed into spots with a diameter of 1.5 cm. The first beam exposed the blackbody to generate PA signals for calibrating laser energy levels changing with laser pulses and wavelengths. A 5 MHz ultrasonic transducer (V302, Immersion Transducers, Olympus Corp., Tokyo, Japan) collected the signals magnified by 35 dB using an amplifier (5073PR, Olympus Corp., Tokyo, Japan). The second laser beam, passing through a hole at the bottom of a small square acrylic container, focused on the cancer with a fluence of 25 mJ/cm$^2$. To minimize the light attenuation caused by water, the distance between the top of cancers and the water surface was maintained at approximately 1 mm. The PA signals from cancers were detected by a needle hydrophone (HNC-1500, ONDA Corp., Sunnyvale, CA) after being increased by 35 dB using an amplifier (5072PR, Olympus Corp., Tokyo, Japan). Both amplified signals were captured by an oscilloscope (HDO6000, Teledyne Lecroy, USA) at a 2500 MHz sampling rate and 32 average times.

### C. Numerical simulations

Light diffusion in tissues was simulated using the Monte Carlo approach in a MATLAB-integrated toolbox called MCmatlab. The PA detection setup defined a 3D square model consisting of air, water, normal tissues, and cancer, as shown in **Figure 1(b)**. Each side of the model had a length of 55 mm, composed of 220 voxels. The malignancy was assumed to be a uniform sphere with diameters of 8 mm, 12 mm, and 16 mm. **Table I** depicts the optical properties of three biomacromolecules and two tissue types at 1580 nm, an absorption peak of collagen. Collagen, water, and lipids are the chromophores showing primary absorbance in the extended NIR window (1200–1700 nm). The $\mu_a$ of tissues were determined using linear spectral mixing:

$$[\mu_a(\lambda)]_i = \sum_{i.molecule} [c_{molecule}]_i \mu_{a,molecule}(\lambda) \quad (1)$$

Where, the variable $i$ represents either normal or malignant tissues, the variable *molecule* interprets collagen, lipids, and water, and $\lambda$ corresponds to a particular wavelength. $c_{molecule}$ and $\mu_{a,molecule}$ indicate the volume fractions and absorption coefficients of three biomacromolecules, respectively. Due to limited availability of data on mouse tissues, we relied on the characteristics of human breast and statistics of our earlier measurements [5]. The water content was considered to be 20% for normal tissues and 9% for malignant tissues, as determined by Shah et al [6]. The values for $c_{collagen}$ and $c_{lipid}$ were defined as 5% and 8% in normal tissues, while 10% and 3% in malignancies. The $\mu_s$ of tissues were calculated from an estimation based on the

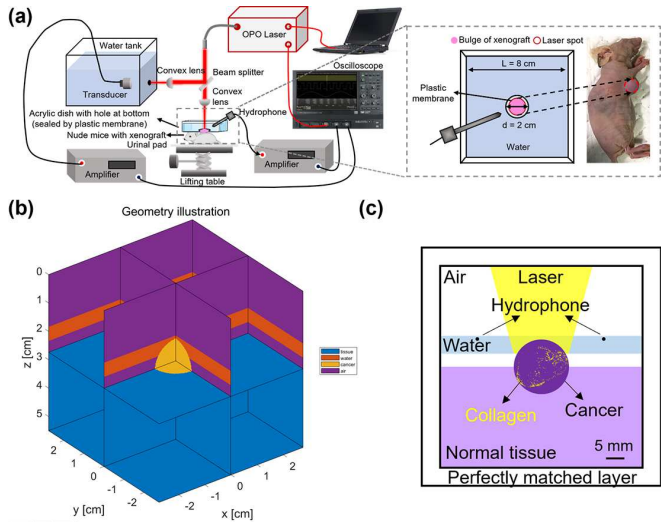

**FIGURE 1**. Photoacoustic (PA) experiment setup and simulation settings. (a) *In vivo* PA detection system. (b) 3D model for light diffusion simulation in MCmatlab. (c) 2D model for PA signal propagation simulation in k-Wave.

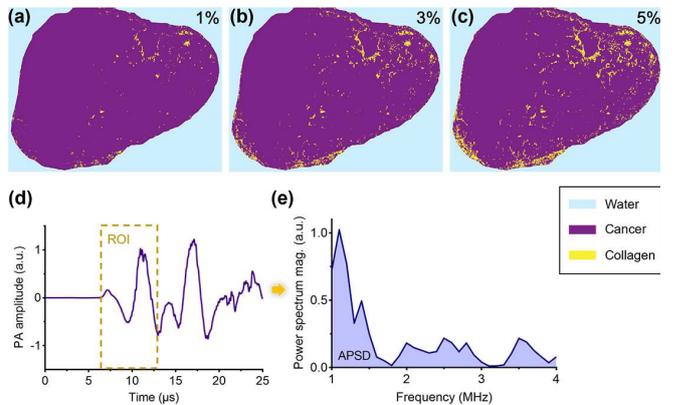

**FIGURE 2**. Numerical simulation samples and PA parameter extraction. (a)-(c) Samples obtained from Masson's trichome staining images of murine breast cancer models, showing relative collagen contents of 1%, 3%, and 5%. (d) Representative PA signals. (e) Representative PA power spectral density and the parameter relative "area of power spectrum density (APSD)". "mag." stands for "magnitude".

Mie's theory:

$$\mu_s = \frac{1}{1-g} a \left(\frac{\lambda}{\lambda_0}\right)^{-b} \quad (2)$$

Where, $g$ denotes anisotropy factor, and $\lambda_0$ is the reference wavelength, equaling to 600 nm. $a$ and $b$ represents the scattering amplitude and power, being extracted from the study by Taroni et al [7]. A vertically incident beam was used as the optical source, directing towards the malignancy and focusing at the bottom of water layer, which had a spot of 15 mm diameter.

TABLE I. OPTICAL PARAMETERS OF TISSUES AT 1580 NM

| Tissue | Absorption coefficient, $\mu_a$ (cm$^{-1}$) | Scattering coefficient, $\mu_s$ (cm$^{-1}$) | Refractive index, n | Anisotropy, g |
|---|---|---|---|---|
| Normal | 1.23 | 74.85 | 1.40 | 0.90 |
| Cancer | 2.49 | 37.73 | 1.40 | 0.85 |
| Water | 9.74 | 3.37 | 1.33 | 0.99 |
| Air | 0 | 0 | 1.00 | 1.00 |
| Collagen | 6.59 | / | / | / |

The production and propagation of PA signals was performed using the k-Wave toolbox, an open-source software in MATLAB 2021b. **Figure 1(c)** depicts the established 2D model through the largest vertical cross-section of cancer. The model had a grid size of 0.125 × 0.125 mm$^2$ and incorporated the collagen distribution based on Masson's trichome staining images. A medical picture processing software, Fiji ImageJ, was utilized to differentiate collagen from other cancer tissues under the guidance of skilled physicians. The collagen growth in cancers was mimicked by adjusting the positive-stained threshold of histological images. The obtained binary images of collagen are displayed in **Figure 2(a)-(c)** and used as simulation samples in k-Wave after being meshed. The relative collagen content, labelled on the top right corner of the figures, was determined by dividing the positive-stained area by the total cross-sectional area of cancer. The initial PA pressure $p_0$ was calculated from the photoenergy deposition using the following equation:

$$p_0(r, \lambda) = \Gamma(r)\mu_a(r, \lambda)F(r, \lambda) \quad (3)$$

Where, $r$ represents a spatial point, $\Gamma$ is the Gruneisen parameter, and $F$ symbolizes the light fluence. We assumed that $\Gamma$ is constant, due to the fact that all of the PA sources studied in this study were water-based, which had close Gruneisen parameters. **Table II** presents the acoustic properties of materials examined in this study, including sound velocity, density, and attenuation coefficients. The ultrasonic attenuation $\alpha$ was determined by the acoustic frequency $f$, according to the power law equation shown below:

$$\alpha = \alpha_0 f^y \quad (4)$$

Where, $\alpha_0$ refers to the prefactor, and $y$ specifies the exponent. Two sensors were positioned approximately 1 mm below the water surface and 1cm apart from the malignancy, replicating the actual conditions in PA measurements.

TABLE II. ACOUSTIC PARAMETERS OF TISSUES

| Tissue | Speed of sound (m/s) | Density (kg/m$^3$) | Power law prefactor, $\alpha_0$ | Power law exponent, y |
|---|---|---|---|---|
| Normal | 1510 | 1020 | 0.75 | 1.5 |
| Cancer | 1510 | 1020 | 0.75 | 1.5 |
| Water | 1414 | 1000 | 0.00 | 0.0 |
| Air | 340 | 1.29 | 1.60 | 1.5 |
| Collagen | 1540 | 1050 | 0.75 | 1.5 |

### D. PA signal processing

In PASA, the collagen content in biological tissue was assessed using a semi-quantitative parameter called relative APSD, which indicated the power of PA signals. The extraction procedures were as follows: First, the PA signal (**Figure 2(d)**) was converted to power spectral density (**Figure 2(e)**) using Welch's approach. Then, the density curve was integrated according to the subsequent equation:

$$\text{Relative } APSD_\lambda = \left(\int_{f_2}^{f_1} p(f) df\right) / A_0 \quad (5)$$

Where, the power spectral density is represented as $p(f)$, and $A_0$ is a standardization coefficient. In simulations, $A_0$ was set to $10^{-4}$, while in experiments, it was $10^5$. To reduce system noises, the integration was carried out within the frequency range of $f_0$ to $f_1$. We established these bonds of integration as between 1 MHz and 4 MHz to align with our previous investigation of murine malignancies [5].

### E. Statistical analysis

The statistical data processing was performed using

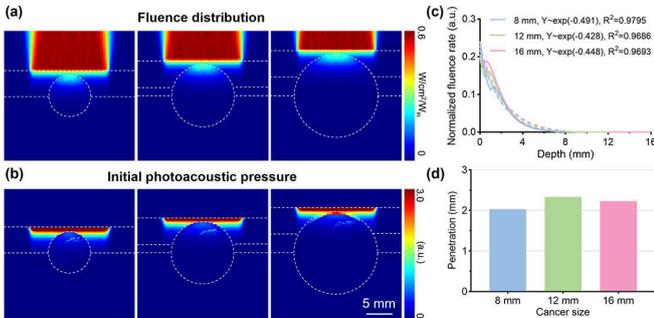

**FIGURE 3**. Numerical simulation results of the optical propagation. Distribution of (a) light fluence and (b) initial PA pressure within the maximum cross-section of cancers. (c) Attenuation of the normalized fluence rate along the central vertical line of cancerous tissues. (d) Penetration depths of light in cancers of different sizes.

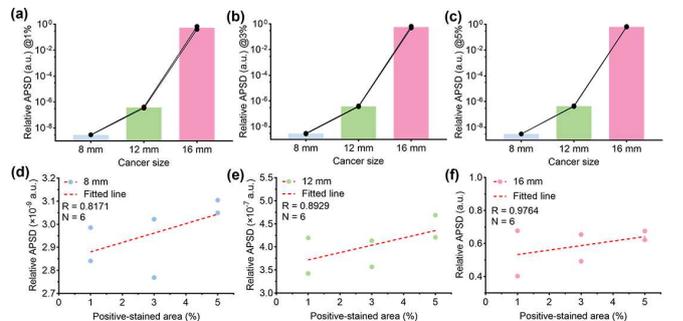

**FIGURE 4**. Numerical simulation results of the PA parameter. (a)-(c) Relative APSDs increase significantly with the growth of cancers owning relative collagen contents of 1%, 3%, and 5%. (d)-(f) Correlation analysis between the relative APSDs and the positive-stained area (%) of Masson's trichome staining images of cancers with diameters of 8 mm, 12 mm, and 16 mm.

GraphPad Prism 9.4.1. Attenuation of the optical fluence rate was determined by fitting to a one-phase exponential decay model. Relationship between the positive-stained area (%) of Masson's trichome staining images and the relative APSDs was analyzed using simple linear regression. The slope R of the fitted line indicated the magnitude of linearity.

## III. RESULTS AND DISCUSSION

### A. Simulation results

**Figure 3(a)** shows the photoenergy distribution inside the vertical plane that passes through the center of malignancies. Observations revealed that light radiation of 1580 nm diffused into cancerous tissues. The optical attenuation graph in **Figure 3(c)** shows an exponential drop in fluence along the middle vertical line of cancers ($R^2 \approx 0.99$). **Figure 3(b)** demonstrates that collagen is the primary PA source within cancer. **Figure 3(d)** depicts the optical penetration depth. The photoenergy of a 1580 nm laser reduced to $e^{-1}$ of its surface value at depths of 2.04, 2.34, and 2.23 mm in cancers owning diameters of 8, 12, and 16 mm, respectively. When the diameters of cancers were almost equal to, but slightly smaller than the light spot, the depth of light penetration in tissues was maximized. The aforementioned findings demonstrated that collagen could produce more pronounced PA pressure compared to other cancerous tissues while exposed to an optical irradiation at 1580 nm. Furthermore, the intensity of initial PA pressure diminished rapidly with increasing depths in tissues.

**Figure 4(a)-(c)** illustrates that as cancer enlarges, there is a noticeable increase in the relative APSDs of samples. Disparities in data from two hydrophones at different orientations were not apparent. **Figure 4(d)-(f)** demonstrate a linear relationship between the relative collagen contents and relative APSDs at invariant cancer size. Moreover, the linearity increased with cancer growing. These findings indicated the APSD is a reliable PA parameter for measuring longitudinal collagen content changes. To improve simulation accuracy, we plan to acquire additional histological slides for modelling and adjust the shapes of malignancies according to their specific aspect ratio. Additionally, we will increase the number of cross-sections for each xenograft to construct a 3D model in k-Wave.

### B. Experiment results

**Figure 5(a)-(c)** depict a representative cancer-bearing mouse at three continuous time points of PA detection, which shows a notable cancer expansion. **Figure 5(d)-(f)** reveal the relative APSDs calculated from *in vivo* PA experiments on three groups of mice grown with varying number ratios of breast cancer cells and CAFs. Despite sample variations, the average relative APSDs exhibited an increasing tendency for the CAFs ($0.0 \times 10^6$) and CAFs ($1.5 \times 10^6$) groups, while showing a contrasting trend for the CAFs ($3.0 \times 10^6$) group. The results suggested that an elevated proportion of CAFs and breast cancer cells might lead to the suppression of collagen production in comparison to the moderate ratios. In future work, extra time points will be set for continuous monitoring. Moreover, histological investigation will be performed at each time point on xenografts to precisely verify the collagen evolution modulated by CAFs.

## IV. CONCLUSION

The purpose of this study was to evaluate the feasibility of using PASA to examine the longitudinal alterations in collagen contents of cancers affected by CAFs. Simulations utilized histological slides and cancer models of progressively enlarged size. *In vivo* PA experiments were implemented on animal models at three consecutive time points after cancers reached average diameters of 8 mm. A quantified PA parameter APSD was computed to characterize the biomacromolecular level. In conclusion, three discoveries were made. First, PASA demonstrated high sensitivity to subtle variations in collagen contents inside malignancies. Second, PASA provided a facile method to evaluate collagen evolution as cancer advanced. Third, CAF quantity played a crucial role in ECM remodeling. The aforementioned results indicated that CAF-targeted therapy with PASA detection has potential for future clinical breast cancer therapeutics.

## ACKNOWLEDGMENT

This project was supported by the National Natural Science Foundation of China (grant numbers 12034015 & 62088101), the Program of Shanghai Academic Research Leader (grant number 21XD1403600), and the Shanghai Municipal Science and Technology Major Project (grant number 2021SHZDZX0100).

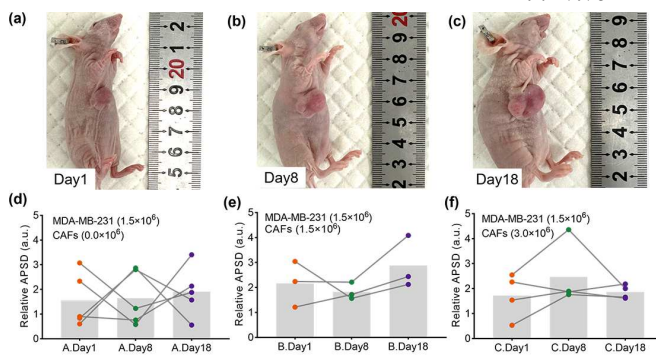

**FIGURE 5.** *In vivo* experiment results. (a)-(c) Representative photographs of breast cancer-bearing mice at three time points for PA detection. (d)-(f) Longitudinal variation of relative APSDs of three murine malignancy groups subjected to different ratios of CAFs and cancer cells.